\documentclass[journal]{IEEEtran}

\usepackage{amssymb}
\usepackage{float}

\usepackage{graphicx}
\graphicspath{{images/}}

\begin{document}

\title{Analyzing Before Solving: Which Parameters Influence Low-Level Surgical Activity Recognition}

\author{Olga Dergachyova, Xavier Morandi, Pierre Jannin
\thanks{Olga Dergachyova, Xavier Morandi and Pierre Jannin are affiliated with INSERM, U1099, Rennes, France and Universit\'e de Rennes 1, LTSI, Rennes, France (email: pierre.jannin@univ-rennes1.fr)}
\thanks{Xavier Morandi is affiliated with Service de Neurochirurgie, CHU Rennes, Rennes, France.}
}%

% \title{Analyzing Before Solving: Which Parameters Influence \\ Low-Level Surgical Activity Recognition}
% \author[label1,label2]{Olga~Dergachyova}
% \author[label1,label2,label3]{Xavier~Morandi}
% \author[label1,label2]{Pierre~Jannin\corref{label_cor}}
% \ead{pierre.jannin@univ-rennes1.fr}
% \address[label1]{INSERM, U1099, Rennes, F-35000, France}
% \address[label2]{Universit\'e de Rennes 1, LTSI, Rennes, F-35000, France}
% \address[label3]{CHU Rennes, Service de Neurochirurgie, Rennes, F-35000, France}
% \cortext[label_cor]{Address: 2 av du Pr. L{\'e}on Bernard, 35043 Rennes, France}

\maketitle

\begin{abstract}
Automatic low-level surgical activity recognition is today well-known technical bottleneck for smart and situation-aware assistance for the operating room of the future. Our study sought to discover which sensors and signals could facilitate this recognition. Low-level surgical activity represents semantic information about a surgical procedure that is usually expressed by the following elements: an action verb, surgical instrument, and operated anatomical structure. We hypothesized that activity recognition does not require sensors for all three elements. We conducted a large-scale study using deep learning on semantic data from 154 operations from four different surgeries. The results demonstrated that the instrument and verb encode similar information, meaning only one needs to be tracked, preferably the instrument. The anatomical structure, however, provides some unique cues, and it is thus crucial to recognize it. For all the studied surgeries, a combination of two elements, always including the structure, proved sufficient to confidently recognize the activities. We also found that in the presence of noise, combining the information about the instrument, structure, and historical context produced better results than a simple composition of all three elements. Several relevant observations about surgical practices were also made in this paper. Such findings provide cues for designing a new generation of operating rooms.
\end{abstract}


% % % % % % % % % % % % % % % % % % % % % % % % % % % % % % % % % % % % % % % % % % % % % % % % % % % % % % % % % % % % % % % % % % % % % % % 

\section{Introduction}

Today, an overwhelming flood of new technologies and equipment threatens to overrun operating rooms, adding even more complexity to the surgical workflow. A large amount of research is focused on smart and situation-aware intra-operative assistance to help alleviate the surgeon's stress and facilitate procedures. Automatic recognition of surgical processes represents a substantial part of this effort. The surgical process can be described in different hierarchical semantic levels: phases, steps and activities \cite{Lalys14}. A surgical phase, the highest granularity level, is a procedure period with several steps and interactions between the surgeon and surgical staff. A surgical step is defined as a sequence of activities with a specific surgical objective. A surgical activity, the lowest semantic level, is a physical activity of the surgeon consisting of the performed action (also called verb), the surgical instrument used, and an anatomical structure on which the action is performed. 
Many research groups have studied the recognition of high-level surgical phases \cite{Forestier15a,Katic16,Twinanda17,Bodenstedt17} and steps \cite{Bardram11,Twinanda15}. A large amount of research has also been devoted to how to recognize surgical gestures \cite{Haro12,Gao16,DiPietro16}, which offer much lower granularity yet no semantics, from training sessions (\textit{e.g.,} on JIGSAWS or MISTIC datasets).
Yet, only a few works exist studying the automatic recognition of low-level semantic activities from real complex clinical procedures \cite{Lalys13,Meissner14}. This automatic recognition is not only useful for in-depth situation awareness, analyzing semantic activities also enables better understanding, learning, and teaching of surgical procedures \cite{Forestier12}. Surgical skills can be objectively evaluated based on a sequence of performed actions \cite{Riffaud10,Forestier13}. Several other applications include detection of deviations from a standard procedure flow \cite{Bouarfa12,Huaulme17}, accurate estimation of remaining time, and resource management \cite{Maktabi17}. 

Due to the lack of automatic recognition, most applications use manually annotated surgical activities, which is a terribly tedious and time-consuming process. However, the automatic recognition of low-level activities is an extremely challenging task. Unlike phases and steps, activities are of shorter duration (seconds \textit{vs.} minutes) and higher diversity in terms of number (hundreds of distinct items \textit{vs.} dozen), execution order (great multitude of possible paths \textit{vs.} simple sequencing) and surgeon/practice-specific characteristics. To facilitate the recognition process, the approaches proposed in the literature break down the activity into its meaningful elements (\textit{e.g.,} verb, instrument, and structure) then proceed with one-by-one detection. The activity is then deduced from one or a combination of elements. The elements to be detected are chosen depending solely on available signals, without any analysis of their relevance. The instrument is often considered a good indicator of the on-going task \cite{Kranzfelder2011,Bouarfa12,Maktabi17}, even though it has been shown to have multiple functions that vary depending on the situation and surgeon \cite{Mehta02}. The verb, which provides pertinent information about the activity context, is difficult to recognize due to a high variability of action execution \cite{Meissner14}, and often requires additional sensors. The anatomical structure, on the other hand, can be recognized from usually available image-based signals \cite{Lalys13}, without extra sensors needing to be brought to the operating room. However, no study as yet exists justifying the choice of elements to detect. 

In this paper, we propose to approach the problem from the opposite direction. Assuming information on all three elements is available, we assess their impact on the performance of low-level activity recognition with the aim of defining a minimum set of required sensors and signals. This is the first large-scale multi-site study of elements' importance to activity recognition, conducted on complex clinical data comprising four different types of surgery. This work's unique contribution consists in its original approach to information analysis for the optimization of operating room sensors, as well as its study results.

\section{Methods}

\subsection{Clinical data}
\label{sec:data}
In order to assess the impact of aforementioned elements on activity recognition, we analyzed four different surgical procedures performed by junior and senior surgeons: anterior cervical discectomy and fusion (ACDF) \cite{Forestier13}, lumbar disc herniation (LDH) \cite{Riffaud10}, pituitary adenoma (PA) \cite{Lalys10}, and cataract surgery (CS) \cite{Lalys13}. The first three neurosurgical procedures were studied via data collected from two university hospitals: Rennes (France) and Leipzig (Germany). Cataract surgery data was taken from the university hospital of Munich (Germany). A total of 154 interventions were studied representing intra- and inter-domain diversity, variety of practices, and a range of skill levels. Table \ref{table:data_info} contains additional information about each of the seven datasets. 

The data consist of manual annotations of the surgical process, \textit{i.e.,} phases and activities. The neurosurgeries were annotated by the same senior surgeon in real time. The cataract surgeries were annotated by the same PhD student based on video recordings. Both annotators had previously been carefully trained to use the annotation software. The first five annotations for each procedure were considered as tests and not taken into account. All annotations were carefully reviewed afterwards using the same software. The phase annotation is a 3-tuple containing its name, start and end time, \textit{e.g.,} \textit{(discectomy, $t_{start}$, $t_{end}$)}. The activity annotation is represented by a 7-tuple consisting of actor, body part, verb, instrument, structure, and start and end times, \textit{e.g.,} \textit{(surgeon, left hand, hold, classic forceps, muscle, $t_{start}$, $t_{end}$)}.

\begin{table*}[t]
\setlength{\tabcolsep}{6.7pt}
\centering
\caption{Information about datasets used for analysis}
\label{table:data_info}
\begin{small}
\begin{tabular}{l|cc|cc|cc|c}
\hline\noalign{\smallskip}
Procedure & \multicolumn{2}{c|}{ACDF} & \multicolumn{2}{c|}{LDH} & \multicolumn{2}{c|}{PA} & \multicolumn{1}{c}{CS} \\
\noalign{\smallskip}
\hline
\noalign{\smallskip}
Hospital				& Leipzig (L) 	& Rennes (R) 	& Leipzig (L) 	& Rennes (R)	& Leipzig (L) 	& Rennes (R) 	& M \\
Number of surgeons 		& 4 			& 5 			& 6 			& 5 			& 2 			& 1 			& 2 \\
Number of interventions	 & 16 		& 48 		& 25 		& 20 		& 15 		& 11 		& 19 \\
Duration (min)		& 156$\pm$61 & 85$\pm$26 & 80$\pm$26 & 34$\pm$14 & 78$\pm$21 & 58$\pm$22 & 12$\pm$3 \\
Number of activities per intervention & 367$\pm$149 & 244$\pm$76 & 242$\pm$72 & 148$\pm$49 & 266$\pm$77 & 213$\pm$46 & 29$\pm$5 \\
Number of unique activities & 377 		& 379 		& 413 		& 243 		& 282 		& 255 		& 45 \\
Number of unique phases 		& 5 			& 5 			& 4 			& 4 			& 5 			& 6 			& 7\\
Number of unique verbs 		& 12 		& 12			& 12			& 11 		& 15			& 15			& 12\\
Number of unique instruments 	& 25 		& 30 		& 27 		& 23 		& 32 		& 30 		& 15 \\
Number of unique structures 	& 11 		& 7	 		& 10 		& 8 			& 7	 		& 9	 		& 7\\
\hline
\end{tabular}
\end{small}
\end{table*}

\subsection{Semantic analysis}
Having information about each surgeon's hands simultaneously is relevant for a more complete understanding of the situation. We thus modified the definition of a low-level activity in this study to include six items: verb, instrument, and structure defined for both left and right hands of the surgeon at the same time.  An example of an ACDF activity could be \emph{(hold, forceps, disc, cut, scalpel, ligament)}. A particular value of an element, \textit{e.g.}, ``cut'' for the verb, is called an \textit{instance}. Using the start and end timestamps, each annotated procedure was then transformed into a temporally ordered sequence of 6-tuple activities. 

In order to assess the impact of each semantic element on the overall recognition process we have used the following scheme. If we want to know how well a whole activity can be recognized solely knowing the used instruments, we can apply a \emph{``010010''} mask to the activity tuple, which gives us \emph{(unknown, forceps, unknown, unknown, scalpel, unknown)} for the example above. The element is considered known for both left and right hands, as in practice the same type of sensor is needed to recognize both. We also take into consideration a temporal context, meaning \emph{n} activities having taken place before. The same mask is also applied to the \emph{n} previous activities, ensuring that only available information is involved in the analysis. The problem then consists of mapping a sequence of partially hidden tuples to a full tuple (\textit{i.e.}, masked with \emph{``111111''}) of a current activity. This problem is resolved using a deep neural network, described in the next subsection. This is performed for all other \textit{configurations}, meaning the solely known elements, as well as their combinations. Finally, the neural network model of each configuration is tested on the activity recognition task, and their performances are compared, as described in Section \ref{sec:study}.

The relevance of the temporal context has previously been discussed in \cite{Forestier15a,Maktabi17}. We found out that working with deep learning and a fairly small dataset requires a careful choice of parameters, \emph{n} in our case, in order to enable an effective learning process. We will talk about our choice of this parameter, defined as a function of factors like dataset size, number of unique activities, average number of activities per intervention, and complexity level, in Section \ref{sec:results}.

\subsection{Deep neural networks for analysis}

Today, deep learning methods are successfully applied to many different problems, starting from image labeling to natural language modeling and text generation. In the majority of cases, they outperform classical machine learning methods in terms of performance. Long Short-Time Memory (LSTM) recurrent neural networks enable analysis of long sequences with complex temporal dependences. We used LSTM in this study, hypothesizing that any hidden elements of an activity depend on currently known elements, as well as on the temporal context. A variant of classic LSTM \cite{Graves12} including three gates (\textit{i.e.}, input, forget, output), an output activation function, no peephole connections, dropouts, and a full gradient training was used. 

While many LSTM models exist, all produce similar state-of-the-art results \cite{Jozefowicz15,Greff17}. We also tested different sets of LSTM parameters, each time varying the number of layers, number of hidden neurons, batch size, learning rate, optimizer, activation and loss functions, and different data representations. All of the tested models generated similar results, with less than 5\% difference. In order to recreate the same analysis conditions for all activity elements, the same LSTM model (that which provided the best results on preliminary tests) was used for all configurations and experiments, described as follows. The model had two stacked temporal layers with dropouts of 0.2, each containing 256 hidden neurons. It was trained during 50 epochs with a learning rate of 0.001 by 128-size batches. Categorical cross entropy was used as the loss propagation function, together with Adam optimizer.

\subsection{Study design}
\label{sec:study}
In order to confirm our hypothesis and determine the essential sensors and signals that are necessary for activity recognition, we conducted a series of several experiments assessing the impact of each element and their combinations on recognition performance, as described below. The performances were compared based on an accuracy score. An activity was considered well recognized when all tuple items were correctly discovered. All the experiments assumed the presence of element information used as input at each moment of surgery. This information was assumed to originate from underlying distinct processing and recognition algorithms, each taking care of its own element. 

Given the relatively small amount of data we had available for deep learning, we performed a full-cross validation for each dataset in a leave-one-intervention-out manner. Moreover, since LSTM uses non-deterministic algorithms for training, we performed three runs for each fold, calculating an average recognition score for three models. 

\paragraph{Experiment 1: One-element configuration}
The first experiment was designed to compare the activity recognition performances achieved  with using each individual element as the only input. We also examined one-to-one relationships between the elements to assess how well one element can be recognized when another another is known. This experiment focused on the sequential aspect only, omitting timestamps and duration of activities. The model thus had only to predict the correct activity tuples in the correct order, without indicating moments of transition. Each recognized activity is supposed to start when all three elements appear in the operating scene (in theory, they have to be recognized at the same time), and end when they disappear, which then implicitly provides the activity duration.

\paragraph{Experiment 2: Two-element configuration}
The second experiment compared combinations of elements, meaning that a pair of known elements was used to infer a complete activity. The same ``no-time'' condition was used, requiring a sequence of activities as output only.

\paragraph{Experiment 3: Activity duration}
Contrary to two previous experiments where the workflow was considered as a sequence of activities only, in this experiment we added the duration of the activity (in seconds) at the end of its input tuple. We then analyzed how knowledge of activity duration impacts inference process. However, the model was still required to predict a 6-tuple only, and no timing was taken into account when computing final accuracy. The constraint of providing duration for input restricts the recognition process, as you have to wait for the on-going activity to finish. This could negatively reflect on on-line applications, yet it is still well adapted to cases where no immediate reaction is needed during the activity, or when only the order of activities is relevant. 

\paragraph{Experiment 4: Noise in input data}
The previous experiments were conducted with the assumption that all the input information was correct. In reality, raw signals coming from sensors may have a certain amount of noise, or some elements may be mislabeled by corresponding recognition algorithms. In this experiment, some element instances in activity tuples were randomly corrupted in order to simulate noise and create more realistic conditions for the analysis. For example, in one input tuple, the value of the right verb ``cut'' could be replaced by another existing verb ``coagulate'', the left instrument ``needle-holders'' by ``classic forceps'' or the right structure ``ligament'' by ``fascia''. In the simulation, the elements, as well as their left and right counterparts, were independently corrupted, meaning that noise occurred at different moments in time for all six items of the tuple. Four types noise were simulated:
 
\begin{itemize}

\item Uniform distribution noise. Using this kind of noise, a corrupted element instance is replaced by another instance of the same element group chosen randomly from a uniform distribution. 

\item Frequency distribution noise. Often, recognition algorithms tend to assign the labels of the most prevalent classes to incorrectly recognized samples. In our case, if an underlying recognition algorithm was trained over the entire procedure, the most commonly represented element instances (in terms of number of samples) would be those that appear in the operational scene for longer than the others. To simulate the behavior of this type of noise, each instance has a chance to be randomly selected proportional to the frequency of its appearance in the dataset computed by duration.

\item Pairwise noise. Another potential occurrence is when the samples of two major classes are mutually mislabeled (\textit{i.e.}, their labels are switched), this is known as pairwise noise.

\item No signal. Sometimes recognition algorithms fail to identify a performed action or an object present in the scene, providing no label at all. In this experiment, a temporal absence of the sensor signal or the algorithm's disability to recognize an element is simulated by simply replacing the corrupted instance with the word ``none''.
\end{itemize}

Three possible configurations were compared in this experiment: 1) input with one knowing element, 2) with two elements, and 3) with all three elements. For the first two configurations, the LSTM models from experiments \#1 and \#2 trained on noise-free data were tested on corrupted data. The third configuration played the role of base line, where all three elements were available with no need of an LSTM model, and the activity was simply defined as their composition.

For all configurations, the noise was simulated at different rates: 5, 10, 15, 20, 25, 50 and 75\% of the corrupted data in the procedure. For instance, for an algorithm recognizing instruments that provides a correct label 95\% of time, 5\% of all instrument instances in the intervention will have a wrong label. The same is applicable to other elements. Given that a correct activity is one where all tuple items are correctly recognized, with 5\% noise for each element, a total amount of corrupted activity tuples may vary from 5 to 10\% for one-element configurations, to 20\% for two-element configurations and up to 30\% for the base line. In the best case scenario, all items in all corrupted tuples have wrong labels, and in the worst case, no more than one item is wrongly labeled in each corrupted tuple. Giving this great variation, noise at each rate was simulated five times, and an average was calculated. As for experiments \#1 and \#2, no time aspect was involved in the analysis.

\paragraph{Experiment 5: Temporal delay}
For the previous experiments, we mostly worked with only the sequencing aspect (with the exception of experiment \#3). No time was taken into account when computing accuracy, which was strictly based on the order and correctness of activities, not on their durations. The elements were supposed to be recognized at the same time as they appeared in the scene. However, the underlying recognition algorithm may experience a certain delay before providing a label. In this experiment, we simulated such a temporal delay. 

As in the third experiment, we first added the duration of each activity to its tuple, and then simulated delay for all available elements. It is important to mention that a change in activity duration is not the only consequence of the delay. In some occasions, this may also cause a shift in activities order by creating new tuples and deleting or altering existing ones. This changes the workflow of the intervention in terms of sequencing and number of activities. The goal of the LSTM model was to discover a sequence of activity tuples without giving their correct durations. However, they were accounted for when calculating final accuracy, which was computed as the sum of durations of all correctly discovered tuples divided by the total duration of all activities in the intervention. Delays of 1, 5, 10, 15, 20, 25 and 30 seconds were simulated. Again, we did not retrain the LSTM models on delayed data and no model was used for the VIS base line.

\section{Results}
\label{sec:results}

\begin{table}[!b]
\setlength{\tabcolsep}{5.5pt}
\centering
\caption{Average activity recognition accuracy (in \%) achieved in experiments 1 and 2. Values in bold indicate the element(s) that provided the best score for each dataset}
\label{table:res}
\begin{tabular}{c|ccccccc}
\hline\noalign{\smallskip}
& ACDF.L & ACDF.R & LDH.L & LDH.R & PA.L & PA.R & CS \\
\noalign{\smallskip}
\hline
\noalign{\smallskip}
V 	& 49.72 & 66.29 & 52.64 & 62.32 & 48.63 & 68.23 & \textbf{92.79} \\
I 	& \textbf{59.08} & \textbf{79.06} & 59.69 & \textbf{74.89} & 58.17 & \textbf{80.25} & 90.40 \\
S 	& 54.50 & 63.27 & \textbf{64.91} & 68.29 & \textbf{60.52} & 60.08 & 90.18 \\
\noalign{\smallskip}
\hline
\noalign{\smallskip}
VI 	& 64.56 & 81.73 & 63.47 & 75.62 & 60.23 & 82.79 & 96.96 \\
VS 	& 84.99 & 83.33 & 91.12 & 90.11 & 85.16 & 87.73 & 97.06 \\
IS 	& \textbf{94.18} & \textbf{96.54} & \textbf{97.30} & \textbf{97.40} & \textbf{97.10} & \textbf{96.81} & \textbf{99.82} \\
\hline
\end{tabular}
\end{table}

\begin{figure}[t]
\centering
\includegraphics[width=0.5\textwidth]{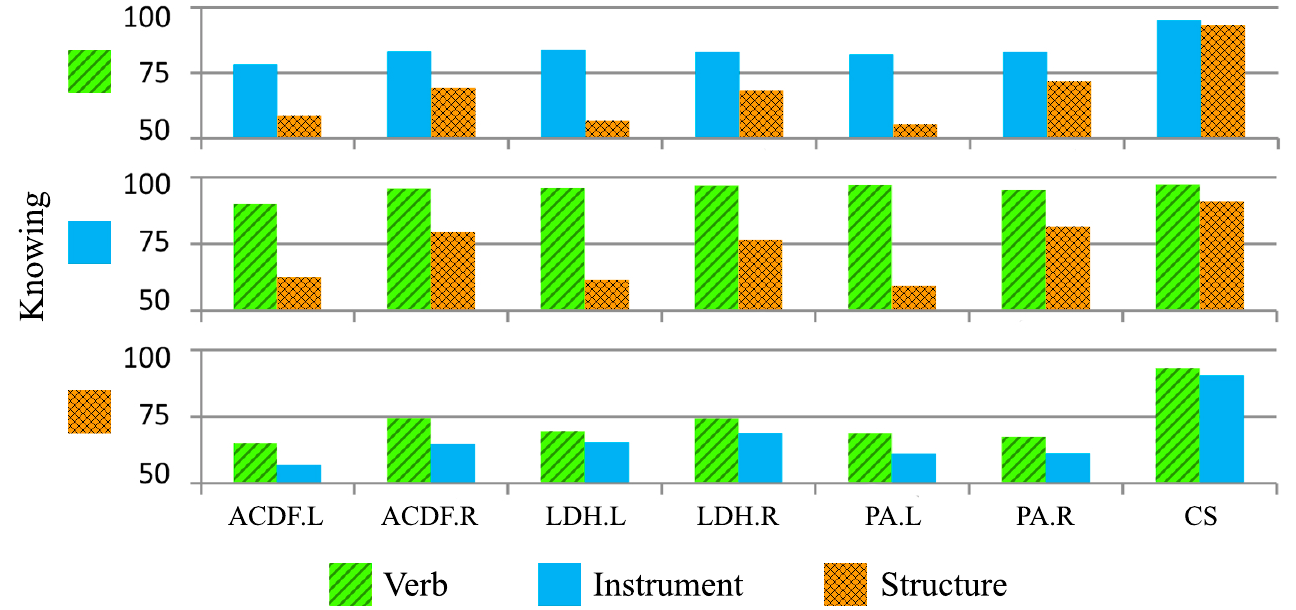}
\caption{Average recognition accuracy (in \%) for one element knowing another}
\label{fig:one_to_others}
\end{figure}

\paragraph{Experiment 1: One-element configuration}
First we assessed the impact of each individual element (\textit{i.e.}, verb V, instrument I, and structure S) on activity recognition performance. The estimated scores are indicated in the upper part of Table \ref{table:res}. The experiment demonstrated that one element is not enough to confidently recognize activity. The instrument provided the best results for four out of seven datasets, yet no element is exclusively preferable for all procedures. The instrument and verb are tightly connected (Fig. \ref{fig:one_to_others}) and have a statistically significant (p-value $\le$ 0.05) correlation, according to Spearman's Rho two-tailed test. While both elements provide a lot of information about each other, they contribute little regarding the structure, and vice versa. Cataract surgery, however, happens to be an exception, it is a short, highly standardized procedure with minimum of deviations and a small number of unique activities. Here, one element almost explicitly defined others, which explains exceptionally high scores.

\begin{figure}[tp]
\centering
\includegraphics[width=0.5\textwidth]{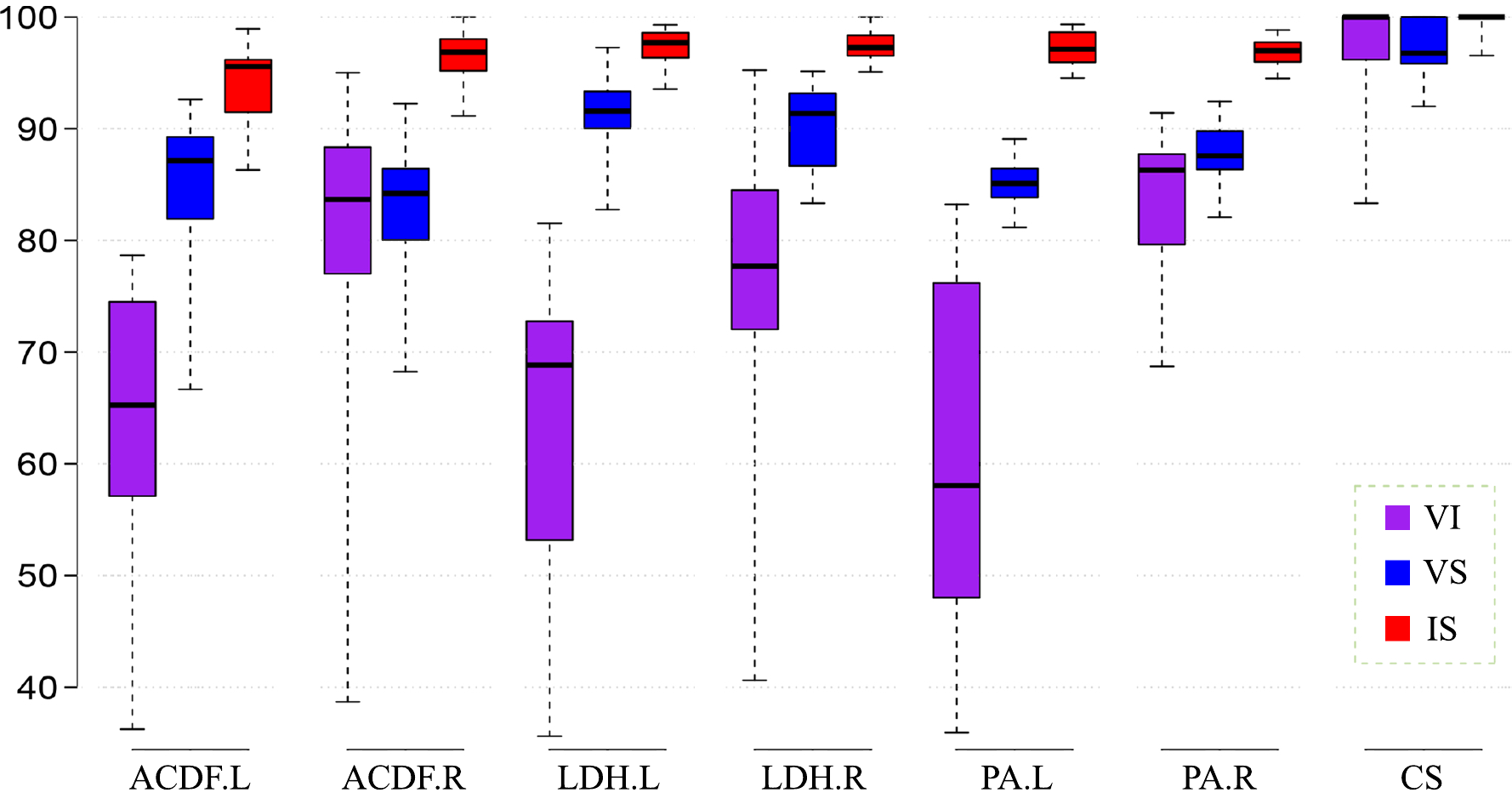}
\caption{Activity recognition accuracy scores (in \%) for element combinations. Center lines of the box plot show the medians, limits indicate the $25^{th}$ and $75^{th}$ percentiles, whiskers extend to minimum and maximum values. For each dataset, VI is on the left, VS in the middle, and IS on the right}
\label{fig:boxplot}
%\vspace{-0.5cm}
\end{figure}

\begin{figure*}[tp]
\centering
\includegraphics[width=1\textwidth]{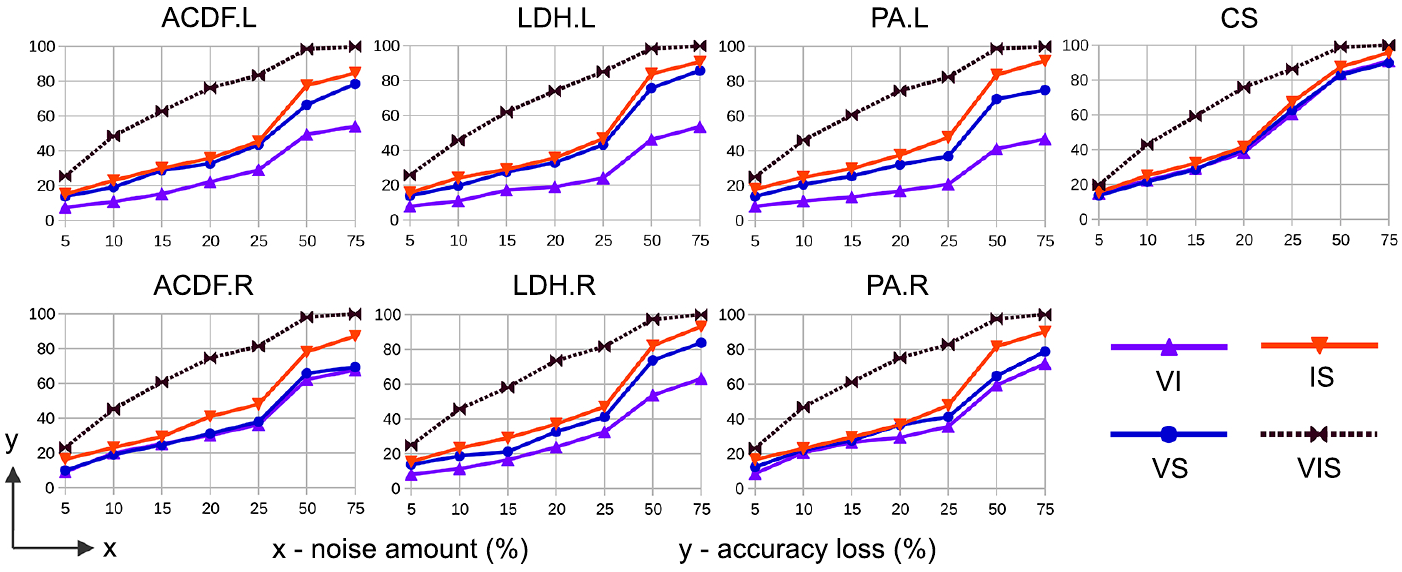}
\caption{Line diagrams showing the loss in activity recognition accuracy with the growth of frequency distribution noise in data for two- and three-element configurations}
\label{losses}
\end{figure*}

\begin{figure*}[tp]
\centering
\includegraphics[width=1\textwidth]{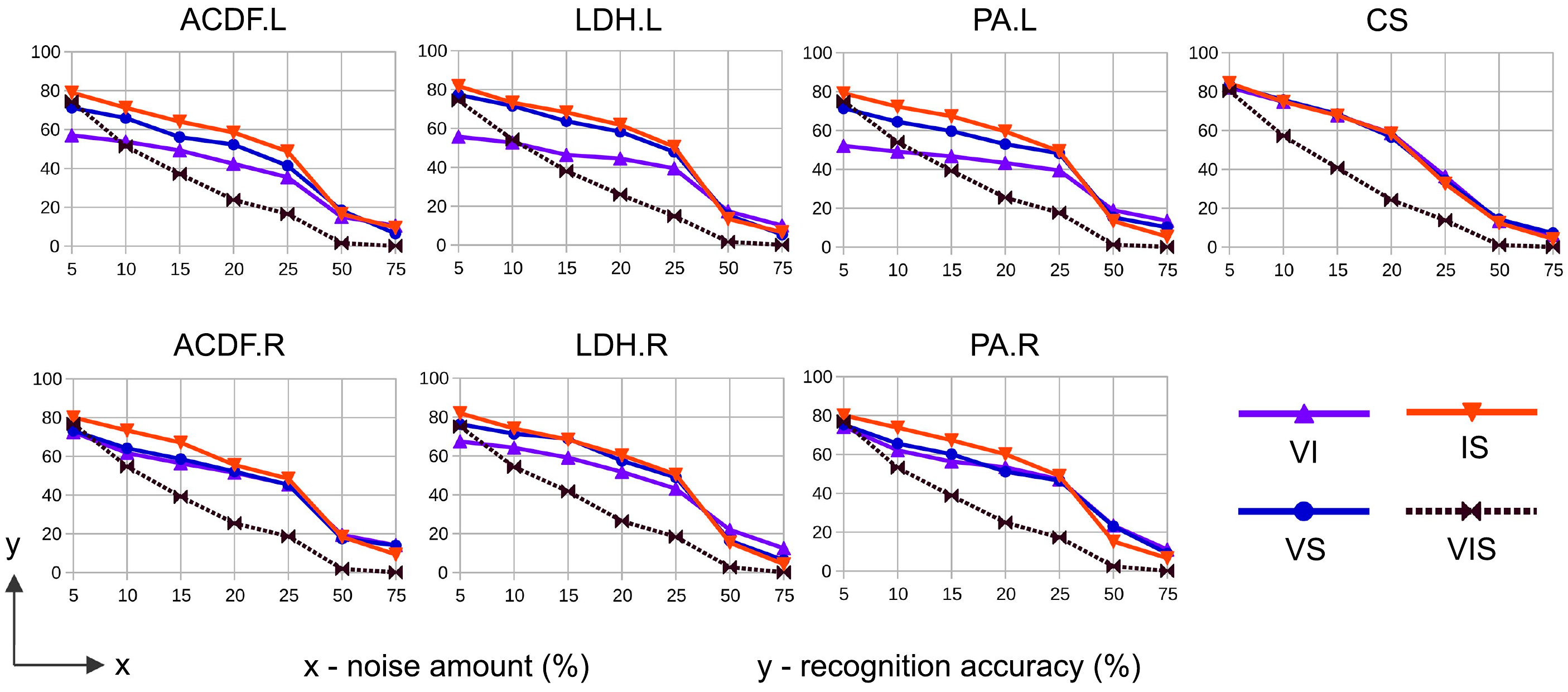}
\caption{Impact of frequency distribution noise on activity recognition accuracy}
\label{frequency}
\end{figure*}

\paragraph{Experiment 2: Two-element configuration}
As one element was clearly not enough to provide satisfactory recognition results,  we conducted an experiment to assess their combinations (\textit{i.e.}, verb-instrument VI, verb-structure VS, instrument-structure IS). The estimated accuracy scores can be found in the lower part of Table \ref{table:res}. We also performed a Wilcoxon signed-rank statistical test. For all datasets, the scores given by these three combinations significantly differed, with a medium to large effect size (p-value $\le$ 0.01 for a two-tailed test, except p-value $\le$ 0.05 for VI versus VS from PA.R and CS, and VI versus IS from CS; no significant difference between VI and VS from ACDF.R). A noticeable difference in scores can be also observed in Fig. \ref{fig:boxplot}. As expected, the VI combination, providing redundant information with less clues about the structure, generated low performance results, proven insufficient for correct stable recognition. A VS combination produced relatively good results of approximately 85\%, that would probably be acceptable for some purposes. For all procedures and sites, the IS combination was satisfactory to confidently recognize activities, producing a score of approximately 95\% and higher. This demonstrates that only two types of sensors are necessary for low-level activity recognition. Far all, the structure, present in two leading combinations, is an essential piece of information.

During the first two experiments, different values of number \emph{n}, defining the size of the temporal context, were tested. We observed that as the temporal window increased, the recognition scores tended to grow quickly until reaching a plateau. With a further augmentation of the number \emph{n}, the recognition performances began to decrease. This behavior is due to the working mechanism of LSTM. In order to obtain a clear picture of the relationship between the elements and activities, the network needs to consider the larger portion of the context. However, in order to clarify these connections, a larger set of training examples is necessary. Its size should increase in correlation with problem's complexity. Without sufficient amount of examples, the learning process becomes much less effective. That is why, calculating the optimal size of the temporal window depends on many aspects and differs for each presented dataset. The best results presented here correspond to $n=50$ for ACDF procedures, $n=20$ for LDH and PA, and $n=5$ for CS.

\paragraph{Experiment 3: Activity duration}
The experiment analyzed the importance of activity duration and revealed that using it as additional input information only slightly improved the activity inference results, while having greater effect on configurations with one known element than those with two. For all datasets, V configuration resulted, on average, in a gain in accuracy of 3.7\%, I - 4.3\%, S - 4.1\%, VI - 2.8\%, VS - 1.3\%, and IS - 1.5\%. Nevertheless, it resulted in the IS combination achieving a recognition accuracy approaching 98-99\%, which corresponds to our hypothesis.

\paragraph{Experiment 4: Noise in input data}

During this experiment, noise was added to input information in order to simulate possible data mislabeling by underlying element recognition algorithms. As expected, all the configurations had reduced ability to predict on-going activity when subjected to noise. Generally, those which previously providing higher recognition scores (\textit{i.e.}, having more useful information in them) were the most significantly affected (see Fig. \ref{losses}). 

While a ranking of one-element configurations was slightly altered for some datasets, the two-element combinations kept their order: IS was still the most informative combination, followed by VS then finally VI. The highest activity recognition accuracy for IS combination ranged from 79\% to 84.4\% with 5\% noise, and decreased to an average of 6.6\% with 75\% noise. However, using an LSTM model encoding procedure history enables the effect of noise to be attenuated as well as ``correcting'' input tuples, especially for smaller amounts of noise. The detailed results for each dataset and noise level can be viewed in Fig. \ref{frequency}. We chose to present the results for the example of frequency distribution noise as this is the most common type of noise related to data recognition and classification.

It is interesting to see that the base line VIS combination always concedes to IS, and that it yields to all two-element combinations event with relatively small amounts of noise (starting from 10-15\% noise). It quickly decreases reaching almost zero accuracy at 75\% noise, and exhibits the most significant accuracy loss (see Fig. \ref{losses}), assuming that a perfect VIS combination would attain 100\% recognition. VIS represents a naive approach of simply putting three elements together with no temporal model, and is thus unable to correct itself. Unlike other configurations, in the presence of noise it is automatically incorrectly recognized. The rapid drop in its performance quality can also be explained by the fact that an additional element in an activity tuple leads to a higher risk of its corruption, especially with greater noise. Thus, having less information is better than having lots with noise.

\begin{figure*}[!tp]
\vspace{-0.5cm}
\centering
\includegraphics[width=1\textwidth]{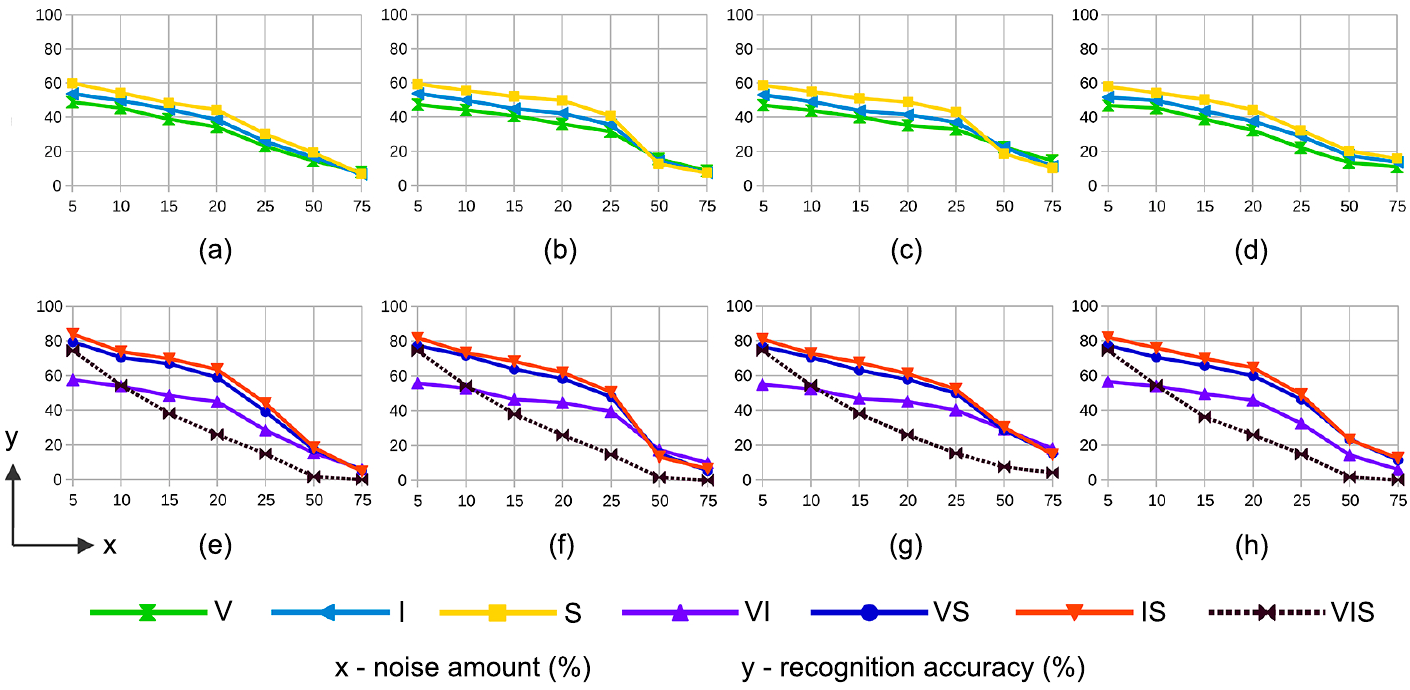}
\caption{Influence of different types of noise on activity recognition performance corresponding to the LDH.L dataset. The uniform distribution noise is present in figures (a) and (e), frequency distribution in (b) and (f), pairwise noise in (c) and (g), no-signal noise in (d) and (h). The top shows the results for one-element configurations, while the bottom for two and three-element ones}
\label{noise}
\end{figure*}

Continuing in our analysis of different types of noise, we found that at lower levels (up to 20\%), the results for all noise types were similar with just a minor difference in accuracy (see Fig. \ref{noise}). This difference grows noticeable at higher noise rates, resulting in steeper or flatter curves. However, no statistically significant correlation between configurations and noise types suitable for all datasets was found at higher noise levels. The curve of the VIS base line is nearly the same for all noise types and datasets. In the case of VIS configuration, the quality of activity recognition on noisy data depends neither on the semantic content of the surgery nor on the type of noise, but rather only on the randomness of corruption. This is evident as an altered tuple item is wrong anyway, no matter its initial or received value.

Previous experiments have demonstrated that within certain limits, a wider temporal window is better for perfectly correct data. This experiment, however, showed that for noisy data, a small \emph{n} value generates better results, as a larger temporal window offers a better chance of making a prediction based on false information. In the exception of one-element configurations of ACDF.L and LDH.L datasets containing uniform noise, $n=5$ is the best option for all other cases. Nevertheless, most of the time, the difference in accuracy scores given by different values of \emph{n} was not statistically significant.

\begin{figure*}[!tp]
\centering
\includegraphics[width=1\textwidth]{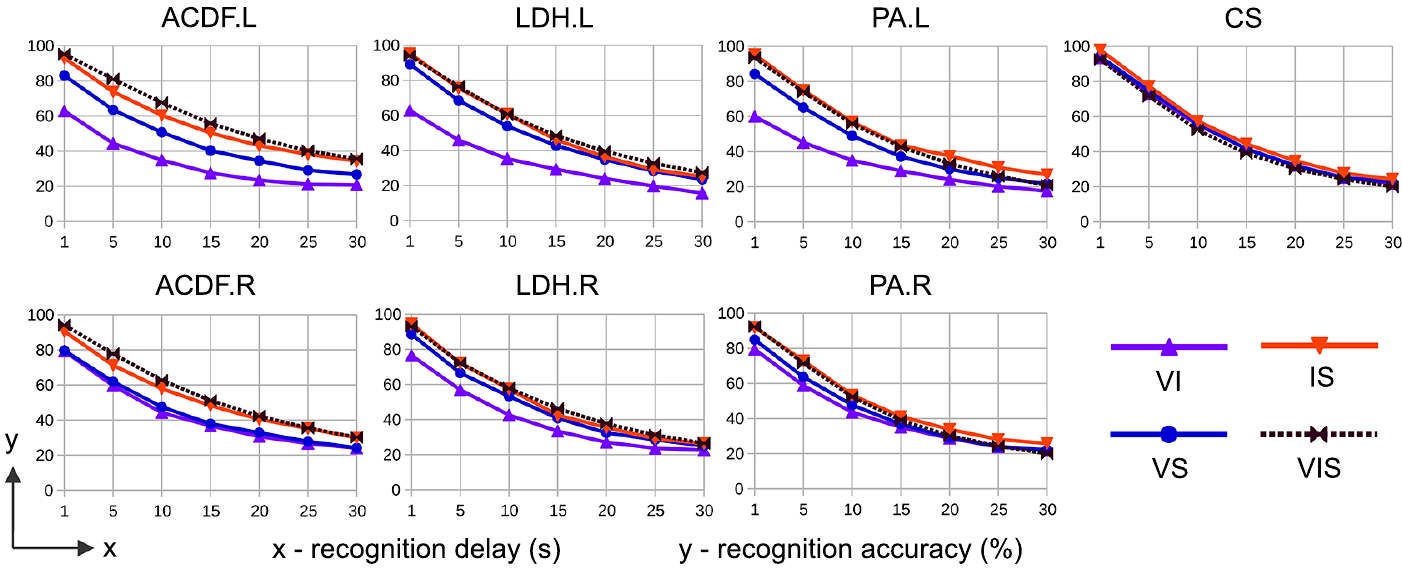}
\caption{Impact of temporal delay on activity recognition accuracy}
\label{delay}
\end{figure*}

\paragraph{Experiment 5: Temporal delay}
This experiment assessed the impact of temporal delay on the quality of activity recognition. As in the previous experiment, we found here that a delay delay caused all the configurations to progressively lose their recognition ability. Nevertheless, as before, the relationship between configurations remained the same with the IS combination achieving the best results. This combination keeps very high scores for a 1s delay, ranging from 91\% to 97.9\%, with an average of 94.1\%. Even if some divergence is observable between IS and VIS curves for several datasets, as seen in Fig. \ref{delay}, the average IS and VIS curves are very similar with less than 1\% difference at each delay point, with the exception of the 30s point where the IS configuration surpasses VIS by 1.8\%.

As was the case with noise, the performance of VIS configuration is progressively impaired as the delay increases, as there is no temporal context of the procedure and no opportunity to correct tuple values. Two-element combinations, on the other hand, we found still able to discover on-going activity due to the history of the procedure represented by the LSTM. They nevertheless suffer from altered activity sequencing, making it difficult for LSTM to follow. We can observe that the biggest deficiencies for two-element combinations occurred in intervals from 1 to 10 seconds (a loss of approximately 15-20\% each time). This can be explained by the fact that during these intervals, the most significant changes in workflows are made (\textit{i.e.}, creation and deletion of activities).

\section{Discussion}
This study proved our hypothesis that for accurate recognition of low-level surgical activity, not all of the activity elements need sensors to track. Though that sort of analysis should also be conducted for other surgical domains, the best choice for neurosurgery is the use of a combination of sensors recognizing the instrument and the anatomical structure. In the case of standardized simple procedures, such as cataract operations where one element is necessarily tightly bounded to two others, searching for the most informative elements is not worthwhile. However, two sensors are sufficient for activity recognition for these procedures as well. The experiments with noise and temporal delay also demonstrated the advantage of the instrument-structure combination over other configurations, including those uniting all three elements, suggesting that the VIS combination can be safely replaced by IS with no significant impairment. However, in order to verify these conclusions, further analysis in certain directions must still be undertaken. First of all, during our fourth experiment, for the sake of simplicity we assumed that all the elements had the same amount of noise in them, which is, of course, not necessarily the case in real-life procedures. The amount and type of noise in each element depends on the underlying algorithm for its recognition. The best way to get realistic estimations of scores is to use confusion matrices from these algorithms to simulate noise in data. We also proceed under the assumption that the perturbation in data was uniform in time, making each time-point equally available for corruption. This aspect should be explored more carefully, as it may not be valid in real surgeries. The same applies to the delay. It may also vary from one element to another in real-life situations, as well as between different element instances. The combinations of different noises and delays must also be evaluated. Secondly, in the last two experiments, conditions under which the studied configurations and VIS base line were compared differed. With the LSTM model, the configurations with one and two known elements benefited from the temporal context of the procedure, enabling input tuples to be corrected when needed. If this was possible for a base line, it would probably generate better results.

We demonstrated that in terms of recognition scores, the IS combination is capable of providing very accurate results. Nevertheless, a considerable drop in performance was observed in the presence of noise and delay (about 80\% accuracy on average at 5\% noise \textit{vs.} 97\% with no noise). This work sought to generate neither a high recognition performance nor a suggestion of an original efficient LSTM architecture. Nevertheless, in order to truly prove that two types of sensor are enough for surgical activity recognition, the overall performance should be enhanced. First, our main focus was on discovering relationships between activity elements using simple LTSM models. There are always subtle connections between the elements that influence the recognition process, however, regardless of which method is used. Thus, the conclusions drawn from the analysis would not considerably change using any other method or LSTM model. However, it should still be possible to find other more suitable deep models that could provide greater accuracy and maintain a strong performance even in the presence of noise. In our experiments, the chosen LSTM model was trained on no-noise data only. One can therefore imagine that retraining the network on simulated noisy data or using some preprocessing methods, as well as noise reduction techniques, could be beneficial. Secondly, the problem with delay can also be avoided. The procedural workflows used in our study were annotated manually in real time. Most of the very short activities are due to an annotator's late reaction or a surgeon's complex hand coordination. For most applications, such an extremely detailed annotation is unnecessary. Eliminating these very brief activities, causing a major change in activity sequencing in the experiment with delay, will make the recognition scores increase again. A larger amount of available data would also provide better results and make the network more robust. In addition, it should be noticed that not all clinical applications require absolute recognition accuracy. Certain errors or delay cause no harm and cab be tolerated, consequently reducing the gap between developing activity recognition techniques and their actual realization and use in operating theaters. Application-dependent metrics similar to \cite{Dergachyova16} may be used to reestimate this gap. Finally, one thing remains clear: placing more sensors in the operating theater is not a solution. The way forward is enhancing underlying algorithms recognizing verbs, instruments and anatomical structures.

In addition to confirming our hypothesis about sensors, this study led to some interesting observations about surgical practices. During the experiments, we noticed that the elements of the right hand of the surgeon were obviously contributing more to the correct identification of the activity. However, despite the correlation between the surgeon's hand movements,  neither the information about the right nor that of the left hand alone was enough to attain acceptable recognition results. This demonstrates how important both hands are in activity execution. The first two experiments also revealed a difference between practices in the Leipzig and Rennes hospitals. Using one-element configurations to discover the activity, the results for the procedures performed in Rennes were always significantly better than for those performed in Leipzig. At the same time, the instrument was clearly a better choice over other individual elements for Rennes, yet the same was not true for Leipzig. Moreover, in all of the procedures conducted in Rennes, we found there was a stronger bound between the instrument and structure, as well as between the verb and structure. Our resulting hypothesis is that, unlike in Rennes, the surgical instruments in Leipzig are more often used for new functions rather than their initially-intended application. This indicates that the procedures performed in Rennes are more standardized and have less variability in surgical workflow. Such observations are important for the analysis and understanding of surgical processes.

\section{Conclusion}
In this work we analyzed the relationships between the essential elements of low-level surgical activity and their impact on  recognition process. By performing a semantic analysis using deep learning, we demonstrated that two out of three elements are enough to confidently recognize an activity. The operated anatomical structure is a crucial element.  The combined structure-instrument pair enables very confident activity recognition, followed by a structure-verb combination that provides slightly worse yet still acceptable results. This knowledge should facilitate the choice of right sensors to install in the operating room of the future for situation awareness. We also made some interesting observations about surgical practices that improve understanding of the surgical process.

%% % % % % % % % % % % % % % % % % % % % % % % % % % % % % % % % % % % % % % % % % % % % % % % % % % % % % % % % % % % % % % % % % % % % % % % 

\section*{Acknowledgements}
This work was partially supported by French state funds managed by the ANR within the Investissements d'Avenir program (Labex CAMI) under the reference ANR-11-LABX-0004.

\bibliographystyle{IEEEtran} 
\bibliography{Sensors}

\end{document}